\relax
\documentclass[letterpaper]{article} 
\usepackage{aaai22}  
\usepackage{times}  
\usepackage{helvet}  
\usepackage{courier}  
\usepackage[hyphens]{url}  
\usepackage{graphicx} 
\usepackage{natbib}  
\usepackage{caption} 
\DeclareCaptionStyle{ruled}{labelfont=normalfont,labelsep=colon,strut=off} 
\usepackage[linesnumbered,ruled,vlined]{algorithm2e}
\usepackage{booktabs}

\usepackage{color}

\usepackage{newfloat}
\usepackage{listings}
\usepackage{tabularx}
\usepackage{amsfonts,amssymb,amsbsy,amsmath}

\usepackage[bookmarks=false,colorlinks=false,pdfborder={0 0 0}]{hyperref}

\title{OVNS: Opportunistic Variable Neighborhood Search for Heaviest Subgraph Problem in Social Networks}

\author{
    Ville P. Saarinen,\textsuperscript{\rm 1,2}
    Ted Hsuan Yun Chen,\textsuperscript{\rm 3}
    Mikko Kivel\"{a}\textsuperscript{\rm 2} \\
}
\affiliations{
    \textsuperscript{\rm 1}Faculty of Social Sciences, University of Helsinki, Helsinki, Finland \\
    \textsuperscript{\rm 2}Department of Computer Science, Aalto University, Espoo, Finland \\
    \textsuperscript{\rm 3}Department of Environmental Science and Policy, George Mason University, Fairfax, USA \\
    ville.saarinen@tuta.io, ted.hsuanyun.chen@gmail.com, mikko.kivela@aalto.fi
}

\begin{document}
\maketitle
\begin{abstract}
    We propose a hybrid heuristic algorithm for solving the Heaviest k-Subgraph Problem in online social networks -- a combinatorial graph optimization problem central to many important applications in weighted social networks, including detection of coordinated behavior, maximizing diversity of a group of users, and detecting social groups. Our approach builds upon an existing metaheuristic framework known as Variable Neighborhood Search and takes advantage of empirical insights about social network structures to derive an improved optimization heuristic.  We conduct benchmarks in both real life social networks as well as synthetic networks and demonstrate that the proposed modifications match and in the majority of cases supersede those of the current state-of-the-art approaches.
\end{abstract}



\section{Introduction}
Identification of dense subgraphs 
has increasingly important applications 
in social networks. Contemporary application areas include detection of structures such as communities \cite{dourisboure2007extraction,chen2010dense}, events \cite{letsios2016finding}, coordinated bot networks \citep{benigni2019bot}, and treatment spillover between individuals in exposure networks \cite{ugander2013graph}. Another interesting development considers using dense subgraph finding for selecting a set of items such that relevant attributes are maximally diversified – a problem that has potential application in selecting a diverse set of users for panels or decision-making bodies \cite{PARRENO2021515}. 

The combinatorial graph optimization problem central to all of these applications is known as the \textit{Heaviest k-Subgraph Problem} (HSP) which involves finding the densest subgraph in a given weighted network. Notably, this problem has been shown to be \textit{NP-hard}, implying that in general it is computationally unfeasible to solve optimally. Particularly in large networks, approximation algorithms and heuristic frameworks are necessary for producing non-optimal yet still high quality solutions to the problem. 

\textit{Metaheuristics} is the area of study that focuses on generalizing heuristic approaches to frameworks that can be applied to any given optimization problem. These approaches are inspired by a diverse set of ideas and disciplines ranging from physics to biology, and chemistry to psychology \cite{hussain2019metaheuristic, sorensen2018history, blum2003metaheuristics}. Generally, metaheuristics have been successfully applied in many areas of science and engineering, and many new methods have been proposed \cite{dokeroglu2019survey}, but
serious limitations have also been recognized, including too much focus on synthetic benchmarking data sets with limited applicability to real world problems  \cite{hussain2019metaheuristic}.


Focusing on developing heuristic algorithms for real-world networks offers a largely unexplored avenue for improving heavy subgraph finding algorithms. More specifically, large social networks exhibit types of structures that could be exploited and targeted by the search algorithms: they are typically sparse networks with large amounts of degree heterogeneity, degree assortativity, clustering, communities, core-peripheries, specific types of link weight embedding, and many other structural features \cite{barabasi1999emergence,newman2003social,barrat2004architecture,onnela2007structure}. Algorithms which take into consideration the large heterogeneity in degrees have been particularly successful in other domains before. For example, degree heterogeneity has been utilized in finding influential substructures such as cores and communities \cite{dourisboure2007extraction} and identification of vital nodes has enabled better understanding about the spreading dynamics that has direct applications in mitigating epidemic outbreaks \cite{lu2016vital, dong2021hunting}. 

Heuristic approaches leveraging heterogeneity in degrees and link weights are particularly interesting candidates for dense subgraph identification algorithms. The distributions of degrees and link weights in social networks have been reported to be close to power-law, log-normal, or other heavy-tailed distributions \cite{barabasi1999emergence, onnela2007structure, barrat2004weighted,clauset2009power}. While the exact shape of the reported distributions have been challenged \cite{clauset2009power,holme2019rare}, for the purpose of finding heavy subgraphs the interest is mostly on the amount of variation in degrees and weights. For example, power-law tails in the range of exponents that are typical of degree and link weight distributions social networks would mean huge amounts of variance that in practical terms grows with the network size. Thus, these features seem promising candidates to consider when identifying members of the most influential subgraphs. Further, as such features are pronounced with network scaling it also makes them especially amenable for designing efficient heuristics for large networks where further reduction in the size of the search space of the given optimization problem is required. 

In this paper we propose a set of improvements to heuristic design, particularly for finding heavy subgraphs in weighted networks, which integrates established heuristic design principles with structural insights derived from empirical network science. More specifically, our approach leverages the heavy-tailed degree distributions characteristic to large social networks and is thus specifically suitable for application areas where networks with heavy-tailed degree distributions, e.g. scale-free networks, is the subject of study. Finally, in order to demonstrate the validity of our approach, we conduct extensive benchmarks in 38 empirical social networks and 41 synthetic networks. Results show that the proposed modifications lead to increased performance against prior variable neighborhood search heuristics as well as more recent state-of-the-art heuristics.

The contribution of our paper is threefold. i) We introduce a heuristic with state-of-the-art performance in large social networks, ii) we demonstrate that insights from empirical network science can be leveraged to improve the design of optimization heuristics in networks, and iii) we produce an efficient and easy to use open source implementation in python, which includes both the original (BVNS) algorithm \cite{brimberg2009variable}, as well as our own improved version (OVNS) of it. By this we aim to make such dense subgraph finding heuristics more readily available and accessible to the broader community of scholars.\footnote{Source code available at \url{https://github.com/Decitizen/OVNS}}


\section{Background}

\subsection{Heavy subgraph finding}
\label{back:optimization}
In this section we introduce the necessary concepts related to finding heavy subgraphs in networks, more specifically in the context of combinatorial optimization problems.

The \textit{heaviest k-subgraph problem} (HSP) is a constrained combinatorial optimization problem
that 
can be defined as: Given a weighted graph $G$ with a weighted adjacency matrix $A$, determine a subset $U \subseteq V$ of size $k$ such that the total edge weight $\sum_{i,j\in U} A_{ij}$ of the subgraph induced by the $U$ on $G$ is maximized. More formally, in HSP the task is to find set $U \subseteq V$ such that

\begin{equation}
\max_{U} \left( \sum_{i,j\in U} A_{ij} \right ) \;\;s.t.\;\;  |U| = k.
\end{equation}

HSP has been studied extensively \cite{billionnet2005different}, and it is closely related to multiple auxiliary graph problems such as the \textit{Densest k-Subgraph Problem} (DSP), which is the special case of HSP where all weights of the graph are either 0 or 1; the \textit{Maximum Diversity Problem} (MDP), another special case of HSP where edge weights are pairwise positive distances such as euclidean distances in $p$-dimensional space; and a well-known combinatorial \textit{Knappsack} optimization problem. Contexts in which solving HSP would be beneficial are manyfold, which explains why the naming convention is rather loose. In the literature it is also known under the names \textit{k-cluster problem}, \textit{maximum edge subgraph problem}, \textit{maxsum problem}, \textit{k-dispersion problem}, and \textit{k-defence-sum problem} \cite{billionnet2005different}.

The decision versions of DSP and its generalized version, HSP, are variants of the original Clique problem that has been shown to be \textit{NP-complete} \cite{karp1972reducibility}. This implies that the optimization versions of the problems are \textit{NP-hard} \cite{corneil1984clustering}. In practice, exact solutions of the optimization problem can be guaranteed to be tractable only for small and sparse graphs, and a small range of $k$ values which is severely restricting for many real-life applications. \citet{letsios2016finding} have proposed a \textit{branch and bound} algorithm which is able to output exact solutions to instances of HSP with $k$ values up to 15 and networks with billions of edges, which translates to order of $[10^4,10^5]$ nodes in dense graphs. However, for solving larger instances one needs to turn to approximate algorithms \cite{asahiro2000greedily,letsios2016finding,brimberg2009variable,marti2013heuristics,hansen2019variable}. Marti et al. (\citeyear{MARTI2022795, marti2013heuristics}) offer two comprehensive surveys to heuristic approaches to solving HSP in the context of MDP.

\subsection{Variable Neighborhood Search}
\label{back:vns}

A variety of metaheuristics have been applied successfully in the context of HSP, including the computationally inspired \textit{Tabu Search}, \textit{Variable Neighborhood Search}, and \textit{Greedy Randomized Adaptive Search Procedure}, as well as \textit{Simulated Annealing}, and \textit{Scatter Search} and \textit{Memetic Search} \cite{MARTI2022795, marti2013heuristics}. 

The Variable Neighborhood Search (VNS) is a metaheuristic framework that was first introduced in \citet{mladenovic1997variable} and has been since successfully applied to both polynomial problems with large polynomial constants and \textit{NP-hard} problems \cite{brimberg2009variable, aringhieri2011comparing, marti2013heuristics, MARTI2022795}. VNS is attractive both because of its conceptual simplicity as well as good general performance in related combinatorial optimization problems in graphs. In the specific domain of HPS and closely related problems such as MDP and DSP, VNS based heuristics have been shown to have extremely good performance superseded only by some of the more recent and more complex heuristics, most notably the \textit{Opposition-based Memetic Search} \cite{zhou2017opposition}. 

The core idea that characterizes VNS is the complementing of the greedy local search phase with a randomized diversification phase. Once the algorithm converges to the local optimum, the search is diversified by replacing portion of the existing solution with randomly selected elements from the set of all elements. For each unsuccessful search attempt, the size of the replaced portion is incrementally increased until all elements in the original solution are replaced. This variability in the size of the perturbation is what gives VNS its name. 

Let $H$ be the solution with highest known objective function value at iteration $t$. VNS then proceeds to iteration $t+1$ by executing the following two steps 
\begin{enumerate}
    \item \textbf{Neighborhood change} (exploration): perturbates $H$ with the aim of finding a solution candidate $H'$ of the search space and thus allows escape from the local optimum of $H$.
    \item \textbf{Neighborhood search} (exploitation): given $H'$ finds a local optimum by greedily exploring the space of possible solutions in the local neighborhood of $H'$.
\end{enumerate}
In the rest of this work, we will refer to this two-step procedure as the \textit{optimization cycle} (also, one iteration of the algorithm). At the end of each optimization cycle, the objective value of the new solution candidate $f(H')$ is evaluated against $f(H)$, and if an improvement is found, $H'$ is adopted as the currently known best solution. Then the next optimization cycle is initiated. This procedure is repeated until a stopping criteria is satisfied. This is typically implemented as either upper bound for execution time or, an upper bound for the number of iterations after the last successful update.

\subsection{Variable Neighborhood Search in the context of networks}
\label{back:vns-hsp}

In the context of combinatorial graph optimization problems such as HSP, both main routines of the VNS are typically implemented as node swapping operations. During the neighborhood search (NS) just one, while in neighborhood change (NC) one or more of the nodes in the current best known solution $H$ are replaced with equal amount of nodes selected from the complement of $H$. More specifically for NS, the replacement node is always selected from the immediate proximity of $H$, while in NC the proximity requirement is relaxed and the domain of selection is extended to cover the whole network. Node selection is typically based on completely random selection, and heuristic approaches aim to exploit network's structural properties or a combination of the two \cite{mladenovic1997variable, brimberg2009variable, aringhieri2011comparing, hansen2019variable}. 

Next, we will introduce some formal notation necessary for describing how VNS operates with respect to HSP. Let $\delta_k(i)$ to be the set of nodes $k$-hops away from node $i$ in the network, and l-neighborhood $\mathcal{N}_l^H$ a set of possible solutions that can be achieved by replacing $l$ nodes in the solution $H$ with $l$ nodes from the set of nodes adjacent to $H$, in other words $\{ \bigcup_{i \in H} \delta_1(i) \} \setminus H$. More formally

\begin{equation}\label{eq:nneighborhood}
    \mathcal{N}_l^H = \left \{ \; H' \;|\; H' = (\, H\setminus B \,) \cup B' \; \right \}\,,
\end{equation}

\noindent where $B \subseteq H$, $B' \subset \{ \bigcup_{i \in H} \delta_1(i) \} \setminus H$ and $|B| = |B'| = l$. As can be observed from the Algorithm \ref{alg:BVNS}, typical VNS starts by enacting a NC procedure in the immediate neighborhood $\mathcal{N}_p^H$ of H  such that size of the perturbation is $p = 1$, and then by each iteration increasing $p$ by $p_{step}$ until either a successful update is found or $p = p_{max}$. At each iteration of the algorithm both NC and NS phases are executed in succession which yields a new solution candidate $H'$. The NS is typically terminated based on a selected improvement strategy; either \textit{first improvement} in which the first improvement over the currently known best solution is returned, or \textit{best improvement} in which the complete $\mathcal{N}_1^H$-neighborhood is explored and the solution with best objective function value is returned. 

In prior work, VNS has been applied to HSP by \citet{brimberg2009variable} with an improved version proposed  by \citet{aringhieri2011comparing}. In a thorough comparison, \citet{marti2013heuristics} compared 315 competing algorithms in various benchmarking data sets. Despite their simplicity, both Brimberg et al.'s BVNS and the modified AVNS variant by Aringhieri et al. achieved performance that was second only to the latest opposition based memetic search by \citet{zhou2017opposition}. 

\begin{algorithm}[!ht]
    \SetKwFunction{NeighborSearch}{NeighborSearch}
    \SetKwFunction{NeighborChange}{NeighborChange}
    \SetKwInOut{Input}{input}\SetKwInOut{Output}{output}
    \DontPrintSemicolon
        
        \Input{$G, H,  p_{min}, p_{max}, t_{max}, p_{step}$}
        \Output{Optimized solution H'} 
        
        \BlankLine
        \While{$t < t_{max}$}{
            $p \leftarrow p_{min}$ \\
            \While{$p \leq p_{max}$}{
                $H' \leftarrow $ \NeighborChange(G, H, p) \\
                $H' \leftarrow $ \NeighborSearch(G, H') \\
                \If{$f(H) < f(H')$}{
                    $H \leftarrow H'$ \\
                    break
                }
                $p \leftarrow p + p_{step}$ \\
            }
            $t \leftarrow t + \Delta t$
        }
    \caption{BVNS algorithm \cite{brimberg2009variable}. Input parameters include the input network $G = (V,E)$ and the initial solution candidate $H \subseteq V$ such that $|H| = k$. $H$ is assumed to have been initialised prior to the execution of BVNS. Other input parameters control the running time $t_{max} \in \mathbb{N}$, maximum size of the perturbation $p_{max} \in [1,\min{(\{k,n-k\})}]$ and the step size by which the perturbation size is incremented, $ p_{step} \in [1,k-1] $.    \label{alg:BVNS}}
\end{algorithm}

\section{Methods}
In this section we describe our modified variant of VNS optimization heuristic.
\subsection{Proposed improvements}
\begin{figure*}[htb]
\centering
\includegraphics[width=1.0\linewidth]{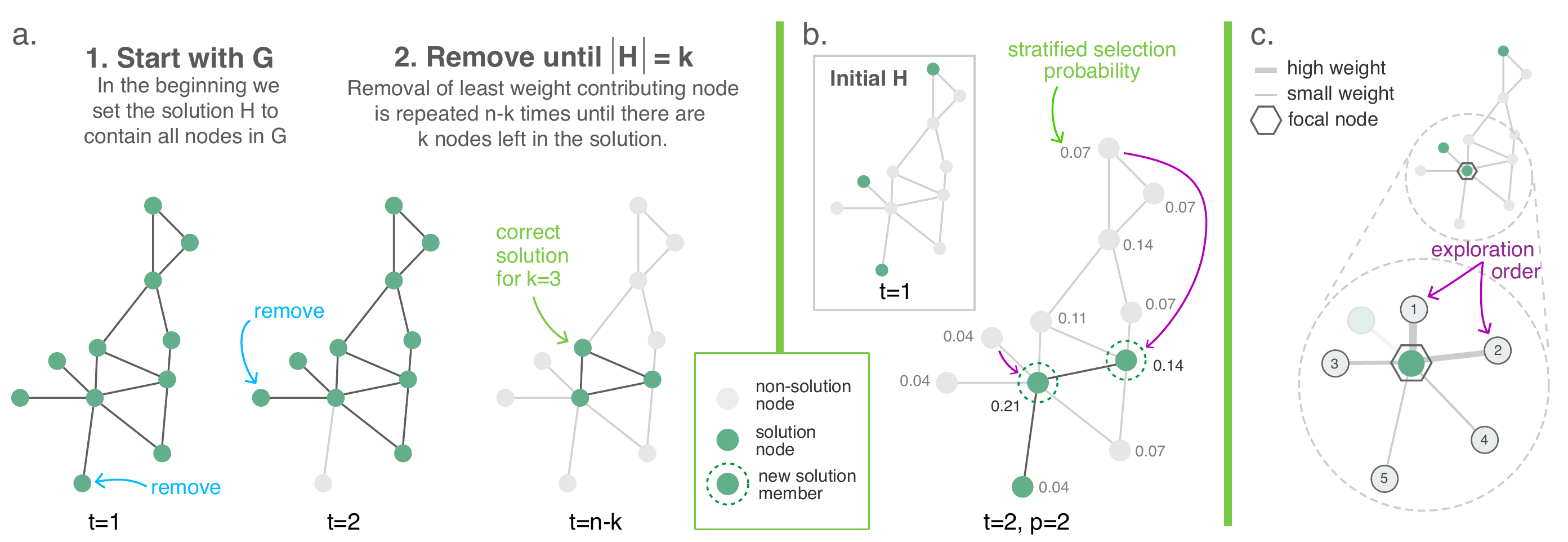}
\caption{Operating principles of OVNS illustrated in a small network. Panel a) shows the drop initialization procedure for three iterations ($t \in \{1, 2, n-k\}$). In small and sparse networks, the drop heuristic is generally likely to find a high quality solution by itself. Panel b) describes the neighborhood change scheme that utilizes stratified sampling with preferential attachment weighting where high degree nodes have selection probability proportional to their degree. Panel c) illustrates the greedy neighborhood search scheme that exploits the heaviest edges in the local neighborhood of the focal node. More specifically, when we iterate through the neighboring nodes of the focal node (hexagon marker), the nodes are explored in descending order based on the weight between them and the focal node, resulting in more efficient neighborhood search and faster convergence to local optima. \label{fig:methods-ovns}}
\end{figure*}

\label{methods:ovns}
In this section, we introduce our main contribution, a modification of BVNS heuristic that we call the \textit{Opportunistic Variable Neighborhood Search} (OVNS). OVNS gains its name by the way it aims to take advantage of the well-established empirical fact that many real-world social networks exhibit heavy-tailed degree distributions which can be explained by the \textit{preferential attachment} mechanism \cite{barabasi1999emergence,toivonen2009comparative}. 
To this end, we introduce following modifications 

\begin{enumerate}
    \item \textbf{Initialization scheme} that constructs the initial solution using the \textit{drop heuristic}.
    \item \textbf{Neighborhood change scheme} that employs the \textit{preferential attachment} style stratified sampling for selecting new nodes into the solution.
    \item \textbf{Neighborhood search scheme} that exploits the heaviest edges in the $\delta_1(i)$ neighborhood.
\end{enumerate}

Figure \ref{fig:methods-ovns} uses a small undirected network to depict the internal workings of the drop heuristic (panel a), a modified NC (panel b), and a modified NS (panel c) schemes implemented in the OVNS.

For the initialization scheme (1) we employ the \textit{drop heuristic} \cite{asahiro2000greedily}. In the drop heuristic, we start with the candidate solution $H$ consisting of all nodes in the graph, remove the node that contributes least to the sum of weights in $H$, after which we repeat the removal $n-k$ times until $|H|=k$ (Figure \ref{fig:methods-ovns}, panel a). For this algorithm, the worst-case approximation ratio for $k=\frac{n}{2}$ has been shown to be bounded to $9/4 \pm \mathcal{O}(1/n)$. See \citet{asahiro2000greedily} for more in depth analysis about other cases. \citet{brimberg2009variable} found empirically that this initialization strategy performs well even without further optimization, which is especially true for small and sparse graphs. 

Next, we improve the neighborhood change scheme (2) by picking replacement nodes from the complement of $H$ and stratifying the picking probability proportional to the weighted degree of the node (Figure \ref{fig:methods-ovns}, panel b). For constructing the sampling distribution, we take advantage of the 
preferential attachment mechanism \cite{barabasi1999emergence} by setting the probability $\pi_u$ of new node $u$ being selected as

\begin{equation}
\pi_u = \frac{s_u}{\sum_{i \in H^C} s_i},
\end{equation}
where $s_i$ is the weighted degree (strength) for node $i$, $H^C$ is the set $V \setminus H$ and $u \in H^C$. 

Finally, we modify the neighborhood search scheme (3) by iterating through the nodes in the solution $H$ and for each $u \in H$ exploring adjacent nodes $v \in \delta_1(u)$ in descending ranking order, where the ranking is based on edge weights $w(u,v)$ (Figure \ref{fig:methods-ovns}, panel c). For accessing the neighboring nodes in ranking order, we produce an arg-sorted copy of the adjacency matrix during initialization of the algorithm, as shown in Algorithm \ref{alg:OVNS} (line 3). 

In large networks, we can further speed up the optimization cycle and reduce convergence time (at the expense of solution quality), by coupling this greedy search strategy with a min-thresholding of the edge weights in the network. To this end, we introduce a quantile-based edge weight control parameter $q = 1 - P(W < w_q)$ where $w_q$ is the threshold corresponding to $q$ value and $W$ is the random variable distributed according to the edge weight distribution of the input network. This is applied at line 2 shown in Algorithm \ref{alg:OVNS}. For example, setting $q=1$ will include the complete set of weights, while $q=0.01$ only includes the top one percent of edge weights. However, evaluation of this optimization step is out of scope for this work and thus we leave it for future inquiry.

\begin{algorithm}[!ht]
\SetKwFunction{DropHeuristic}{DropHeuristic}
\SetKwFunction{EdgeWeightThreshold}{EdgeWeightThreshold}
\SetKwFunction{ArgSortRows}{ArgSortRows}
\SetKwFunction{NeighborSearch}{NeighborSearch}
\SetKwFunction{NeighborChange}{NeighborChange}
\SetKwInOut{Input}{input}\SetKwInOut{Output}{output}
\DontPrintSemicolon
    
    \Input{$A, S, k, q, p_{\text{min}}, p_{\text{max}}, p_{\text{step}}, t_{\text{max}}$}
    \Output{Optimized solution $H$}
    \BlankLine
    $H \, \leftarrow $ \DropHeuristic($A$, $k$) \\
    $A' \leftarrow $ \EdgeWeightThreshold($A$, $q$) \\
    $A' \leftarrow $ \ArgSortRows($A'$) \\
    \BlankLine
    \While{$t < t_{\text{max}}$}{
        $p \leftarrow p_{\text{min}}$ \\
        \While{$p \leq p_{\text{max}} \land t < t_{\text{max}}$}{
            $H' \leftarrow $ \NeighborChange(A, k, H, S, p) \\
            $H' \leftarrow $ \NeighborSearch(A', k, H', S) \\
            \If{$f(H') > f(H)$}{
                $H \leftarrow H'$ \\
                break
            }
            $p \leftarrow p + p_{step}$ \\
            $t \leftarrow t + \Delta t$ \\
        }
    }
    \caption{High level description of OVNS. During the initialization, the drop heuristic is applied (line 1), the network is thresholded according to thresholding parameter $q$ (line 2), and indexing matrix $A'$ is created by arg-sorting rows of the adjacency matrix (line 3). After initialization procedure (lines 1-3) execution enters the outer loop (lines 4-13) which handles the gradual increase in the size of neighborhood change while the inner loop (lines 6-11) executes the optimization cycle. Size of the perturbation is dependent on the number of consecutive unsuccessful optimization cycles (line 12) and it is reset to minimum value (line 5) after execution of the inner loop. \label{alg:OVNS}}
\end{algorithm}

\subsection{Benchmarks}
In order to demonstrate effectiveness of the OVNS, we run it with varied settings against two other HSP heuristics developed in the \textit{Operations Research} literature for the \textit{Maximum diversity problem} (MDP). Algorithms are the \textit{Variable neighborhood search} (BVNS) developed by \citet{brimberg2009variable} and one of the more recently introduced top performing MDP-heuristics \textit{Opposition Based Memetic Search} (OBMA) by \citet{zhou2017opposition}. In order to be able to measure performance of each approach independent of implementation details, we implemented all algorithms in python. Further, to speed up the implementations, we JIT compiled the code base using the Numba library\footnote{Source code for the implementation that uses python and numba libraries is available at \url{https://github.com/Decitizen/OVNS}}. All benchmarks were run in the same computing environment. 

First, we run the three algorithms in a set of 38 social networks with the intention of measuring the general performance of our approach (see Appendix \ref{app:data-sources} for data description). We bound the size of these networks to range $[10^3,10^5]$ nodes. In order to measure the performance across different difficulty levels we vary the size of the targeted subgraph as $k \in \{125, 250, 500, 1000, 2000\}$. 

Second, to demonstrate performance in relatively large and dense networks, we use 5 transformed networks – two $10^4$ node networks which have been generated using a weighted preferential attachment model \cite{barrat2004weighted}, and three differently sized real life networks that are based on Twitter retweet data \cite{chen2021polarization}. Before running the benchmark runs, we transform the networks into dense networks using the non-backtracking version of the Katz communicability \cite{arrigo2022weighted}.

Finally, we run the algorithms in 41 synthetic benchmarking instances from the mdplib2.0 library \cite{marti2021mdplib}. These networks are commonly used in the MDP heuristics literature for comparing heuristic performance and should therefore give a standard baseline of OVNS performance. In the mdplib2.0 library, each network is accompanied by a single predetermined $k$ value. Note that in contrast to many social networks with heavy-tailed edge weight and degree distributions, in MDP networks the edge weights are typically normally distributed.

In general, each algorithm is run 20 times per combination of $k$ value and network, with each run taking 10 minutes. However, for larger and denser instances we accommodated for longer convergence times by finetuning the parameter choices further. For the largest social network ($N=33,696$) we extended the range of $k$ values to cover $\{4000,8000,10000,12000\}$ . Additionally, to ensure satisfying degree of convergence, we redistributed the computing resources for this network such that we ran 15 runs with each run having 30 minute time budget. Similarly for the dense social networks, we run 20 runs, each run having 60 minute budget (see Appendix \ref{app:hyperparam-selection} for more details).

For all three benchmarking data sets, we compute both average and median relative deviations for each algorithm. Similarly, we compute ranking in terms of objective function values such that for each combination of network and parameter setting, the runs are combined to a pool and ranked. Finally, we combine all runs at the data set level and compute median and mean ranks for each algorithm. We define relative deviation as $ d(f^*,f_{H}^A)= 100 \% (f^* - f_{H}^A)/f^{*}$, where $f^*$ is the best objective function value for the given combination of network and parameter $k$ value, and $f_{H}^A$ is the objective function value for the given solution $H$ produced by algorithm $A$.

\section{Results}

\begin{figure}
    \centering
    \includegraphics[width=0.8\linewidth]{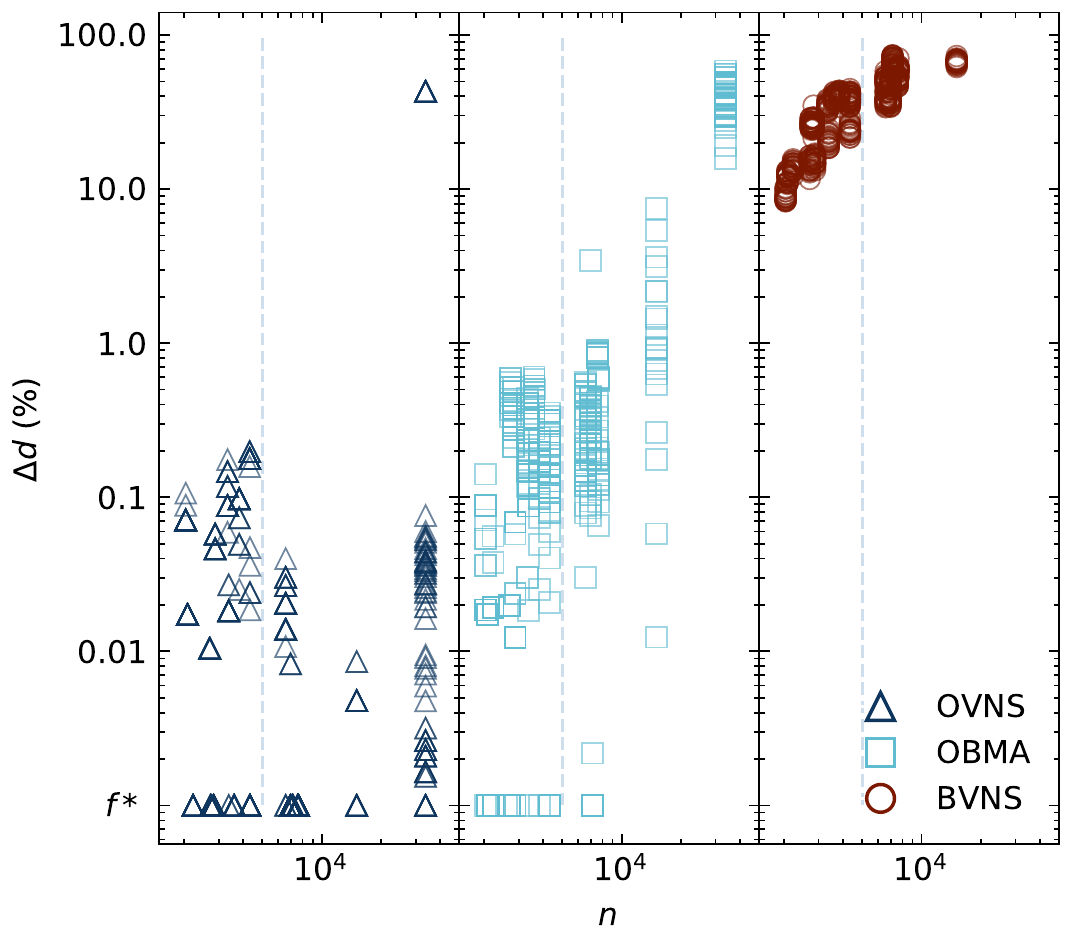}

    \includegraphics[width=0.8\linewidth]{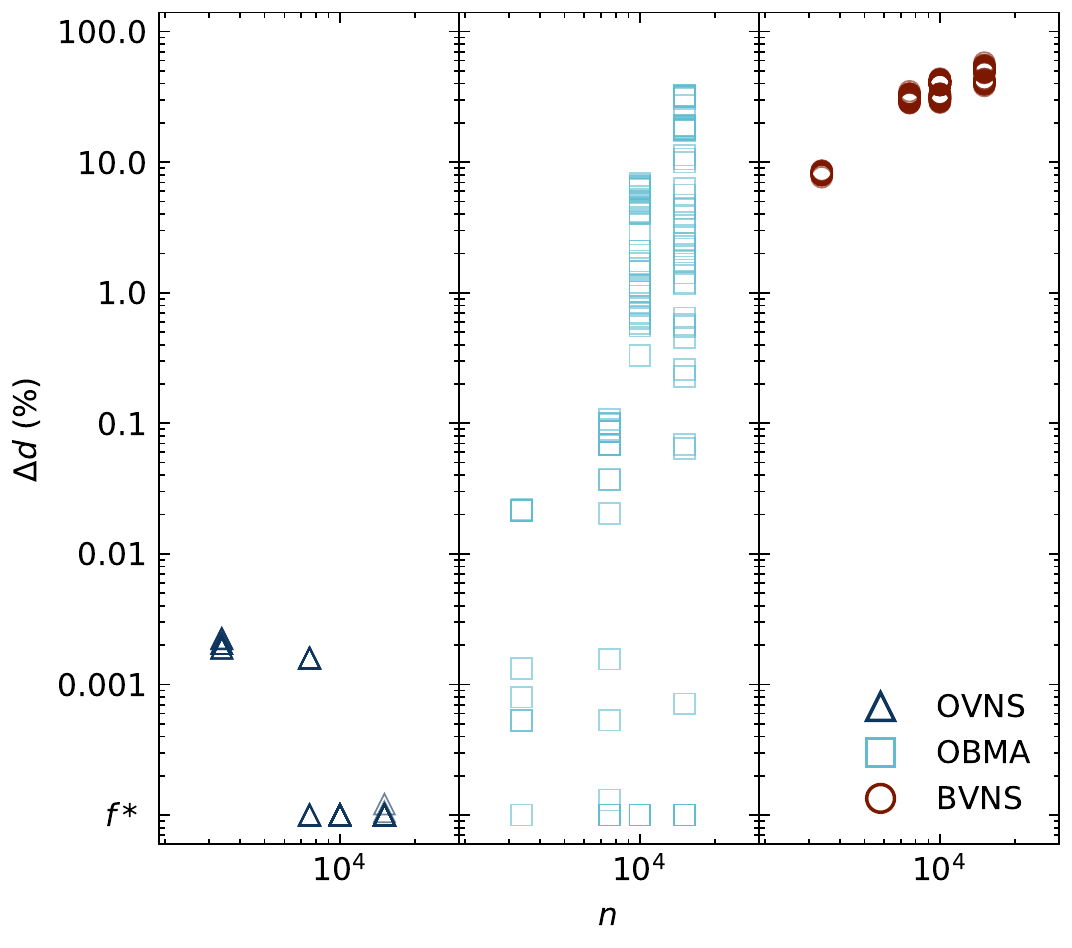}

    \includegraphics[width=0.8\linewidth]{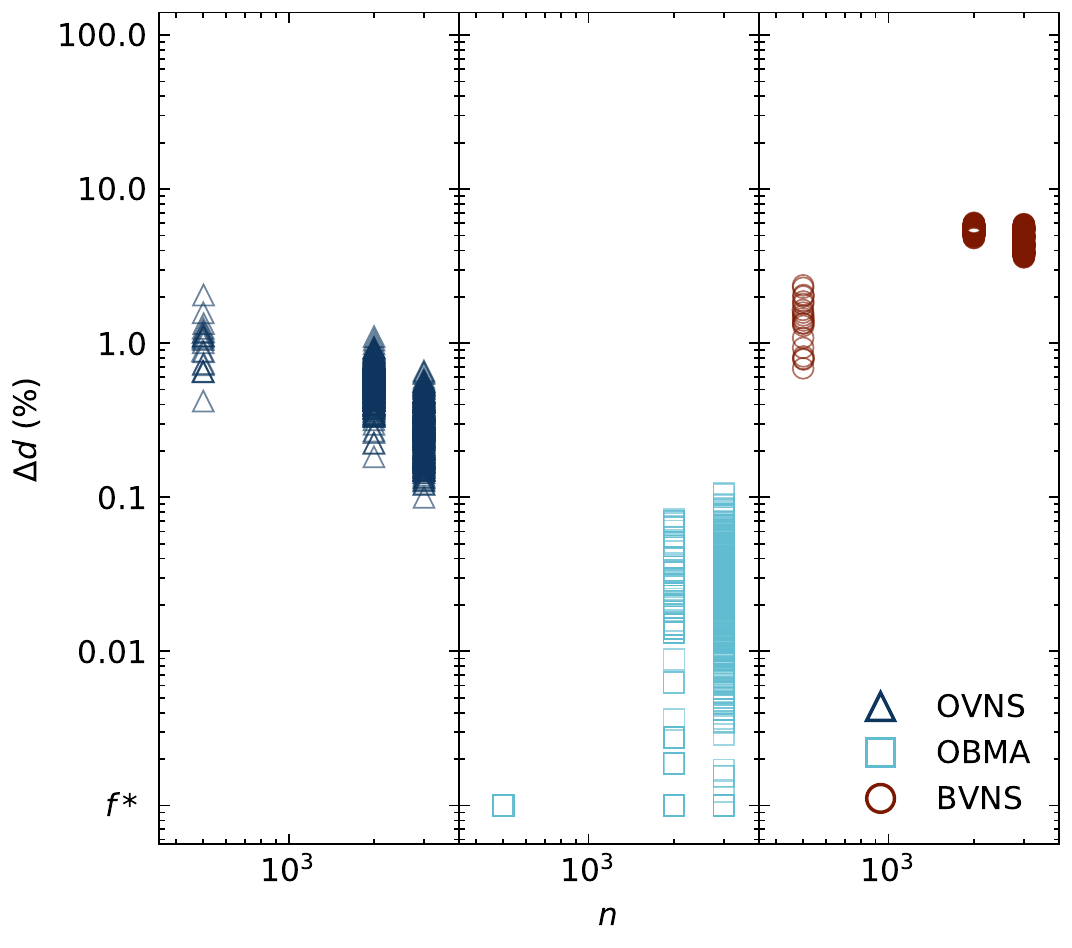}
   
   \caption{Benchmark results in sparse social networks (top), dense social networks (middle) and synthetic MDP networks (bottom). Relative deviations ($\Delta d$) are shown as a function of network size $n$. Smaller values indicate better performance. In small sparse networks OBMA and OVNS demonstrate similar performance, while for $n > 5000$, performances diverge and OVNS consistently outperforms OBMA. Differences are further pronounced in dense networks while in MDP networks OBMA outperforms OVNS. BVNS has worst performance across all network types and sizes, but has smallest deviation in MDP networks. \label{fig:ovns_benchmarks}}
    
\end{figure}

\begin{table}[!htb]
    \centering
    \begin{tabular*}{\columnwidth}{lrrrr}
        \midrule
         OVNS   &  $\bar \mu_r $ &  $M_r$ & $\bar \mu_d $ & $ M_d$  
         \\
        \midrule
        Sparse ($N=20$) & \textbf{2.44} & \textbf{2.0} & \textbf{0.97} & \textbf{0.02} \\
        Dense \hspace{0.05cm}($N=5$) & \textbf{6.10} & \textbf{5.0} & \textbf{0.00} & \textbf{0.00} \\
        MDP \hspace{0.1cm} ($N=41$)  & 21.25 & 21.00 & 0.47 & 0.44 \\
        \midrule
         OBMA &  \multicolumn{4}{r}{} \hspace{0.1cm} 
         \\
        \midrule
        Sparse ($N=20$)  & \hspace{0.05cm} 6.94 & 6.00 & 7.15 & 0.2 \\
        Dense \hspace{0.05cm}($N=5$)    & \hspace{0.05cm} 15.12 & 15.00 & 2.86 & 0.02 \\
        MDP \hspace{0.1cm} ($N=41$)  & \hspace{0.05cm} \textbf{5.49} & \textbf{3.0} & \textbf{0.02} & \textbf{0.01} \\
        \midrule
         BVNS &  \multicolumn{4}{r}{} \hspace{0.1cm} 
         \\
        \midrule
        Sparse ($N=19$)  & \hspace{0.05cm} 22.36 & 21.0 & 36.47 & 37.21  \\
        Dense \hspace{0.05cm}($N=5$)  & \hspace{0.05cm} 33.83 & 34.0 & 36.84 & 39.77  \\
        MDP \hspace{0.1cm} ($N=41$) & \hspace{0.05cm} 40.72 & 41.0 & 4.92 & 5.26  \\
    \bottomrule
    \end{tabular*}
    
    \caption{Table listing the results for the three benchmarking data sets; both sparse and dense social networks (results are subsetted $k \geq 1000$), as well as synthetic MDP networks. For each combination of algorithm and benchmarking data set, means ($\bar \mu $) and medians ($ M $) are reported for both ranking ($r$) and relative deviation ($d$). Measured value of best performing algorithm for each data set is highlighted by bolding. Overall, OVNS demonstrates best performance in 8 out of 12 measures, clearly excelling in large and dense social networks, while OBMA achieves the best performance in MDP networks. 
    \label{tab:ovns_benchmarks}}
\end{table}
\label{chapter:results}

\subsection{Results on small $k$ values}

In both sparse and dense social networks, for $k < 1000$ values OBMA and OVNS find exactly the same optima. This is indicated by median deviations that approach 0 and suggests low problem difficulty which is further corroborated by the fact that for $k < 10^3$, OVNS converges in a fraction of the available time. For OVNS convergence to final value takes on average $37 \pm  108$ seconds in sparse and $266 \pm 515$ seconds in dense social networks (large variance is explained by the large variation in network sizes). In order to find demonstrable differences in performance between OBMA and OVNS, we focus our attention on large problem instances with the additional constraint that $10^3 \leq k \leq n/2$. This upper bound ensures that OBMA has comparable performance to the results obtained in \citet{zhou2017opposition} where the opposition based search was not implemented for $k > n/2$. With these constraints, the number of sparse networks considered in the final analysis is reduced to $20$.

\subsection{Overall OVNS performance}
Table \ref{tab:ovns_benchmarks} depicts the median and average deviations as well as median and average rankings for each algorithm in each of the three benchmarking data sets while Figure \ref{fig:ovns_benchmarks} shows the relative deviations as a function of network size. Overall, OVNS demonstrates the best overall performance; it scores best in 2 of the 3 benchmarking data sets, stays within 1\% mean relative deviance in all data sets, and scores best in 8 out of the 12 aggregate measures across different data sets. We can observe that it clearly excels in dense social networks, followed by sparse social networks. As expected, in MDP networks OBMA is able to outperform it achieving best results in all related 4 aggregate measures.

From Figure \ref{fig:ovns_benchmarks} we can observe the effect of network size to the performance profiles of the algorithms. In sparse networks, once the network size grows beyond $n > 5000$ OBMA's performance starts to deteriorate leading to a visible 1-2 order of magnitude difference in relative deviation between the two algorithms. A similar but even more pronounced trend can be observed in dense networks where OVNS consistently achieves best solutions with both mean and median relative deviation approaching zero while for OBMA the corresponding measures are 2.86\% and 0.02\% respectively. The distributions for the two algorithms differ significantly both in sparse and dense networks. Corresponding significance tests, in sparse (Wilcoxon-Mann–Whitney $U = 11,751$, $n_1 = n_2 = 240$, $\text{p-value} \ll 0.05$, one-tailed) and dense (Wilcoxon-Mann–Whitney $U = 52,768$, $n_1 = n_2 = 550$, $\text{p-value} \ll 0.05$, one-tailed). However, in the MDP benchmarking instances OBMA still retains its performance edge achieving mean and median relative deviations of 0.02\% and 0.01\%, compared to 0.47\%, and 0.44\% for OVNS (Wilcoxon-Mann–Whitney $U = 672,398$, $n_1 = n_2 = 820$, $\text{p-value} \ll 0.05$, one-tailed). The difference between BVNS and OVNS performance is also lower here which further suggests that in these network OVNS does not benefit from the stratified sampling to the same degree as in the two earlier data sets.

\subsection{Improved performance over BVNS}
In terms of relative deviance, BVNS results range between 20-40\% in social networks, but decrease to around 5\% in MDP networks. Somewhat unexpectedly, BVNS instances struggle to converge within the given time limits already at $k < 10^3$ in sparse social networks. Further diagnostics shows difficulties in keeping up in terms of iteration speeds – the number of iterations at the end of each run is on average only $6\%$ $(1.5\cdot10^5)$ relative to OVNS. 

To estimate the performance difference independent of time limit, we conducted 30 additional runs of both OVNS and BVNS algorithms with a budget constraint of 50,000 iterations in one of the sparse Twitter networks (``Left 3'', $N=1226$). In order to control for the effect of the drop initialization, we also ran a modified version of OVNS which uses BVNS's randomized initialization routine. 

Here again controlling for time, regular OVNS achieves the best performance with mean relative deviation of $0.02 \% \pm 0.02$, while for BVNS the corresponding relative deviation is $8.17 \% \pm 0.82$ (Wilcoxon-Mann–Whitney $U=0.0$, $n_1 = n_2 = 30$, $\text{p-value} \ll 0.05$, one-tailed). The main finding is that, when we control for the iterations speeds, the BVNS performance difference drops from the 20-40\% range to below 10\%, yet it still remains significant. If we further control for the effect of the initialization by the drop heuristic, OVNS scores $ 0.08 \pm 0.06$ which is only $0.06$ pp, yet still statistically significant difference (Wilcoxon-Mann–Whitney  $U=198$, $n_1 = n_2 = 30$, $\text{p-value} \ll 0.05$, one-tailed), to the OVNS initialized with drop heuristic. This reveals a secondary empirical finding; the effect of drop initialization to the performance of OVNS is negligible when we compare it to the difference in relative deviations between BVNS and regular OVNS. When combined these findings suggest that the updated NC and NS routines produce significant performance improvements over BVNS. 

\section{Discussion}
\subsection{Conclusions}
In this paper we have proposed an improved version of VNS metaheuristic for subgraph finding, and we have shown that these improvements result in significant performance gains. Particularly in networks with log-normal to heavy-tailed degree and edge weight distributions, OVNS exhibits significant improvement in performance against the original BVNS algorithm, and more importantly, it supersedes the performance of current state-of-the-art heuristic algorithms. OVNS achieves this by both systematically exploiting heaviest ties in the local neighborhood, and by successfully applying what \citet{hussain2019metaheuristic} describe as \textit{intelligent sampling}, i.e. a stratified sampling method is used to restrict or guide the search in the problem space. For this, OVNS incorporates key structural insights from empirical studies of networks, namely by preferentially targeting the heterogeneity in the degree structures of the networks. However, the benefit of preferential sampling only applies where such structures exist, which is demonstrated by the benchmarks in the MDP networks. In these networks OBMA is still able to produce slightly better results, though the margins are small.

\subsection{Limitations}
Although we expected improvement over BVNS performance, the observed difference in performance in favor of OVNS was significantly high. We showed that it is likely that large parts of this difference is explained by the decreased iteration rates of BVNS. This is somewhat unexpected since both OVNS and BVNS use partially overlapping source code and call exactly the same functions at the implementation level. This might suggest that the OVNS benefits from additional compiler or runtime memory optimizations. Specifically, the neighborhood search improvements and the related arg-sorted indexing matrix can be exploited for fast caching during execution (only the most frequently accessed neighbors need to be kept in the memory) while BVNS does not have this feature. However, we also showed that even if we control for the iteration rates, OVNS retains significant 
performance edge over BVNS. The generalizability of this finding is to be taken with caution, as the performance was evaluated in only a single network.

We should also mention that two of the dense networks were created using the weighted preferential attachment model for which the generation mechanism is almost identical to the updated stratified sampling implemented in OVNS. It is therefore to be expected that OVNS performs well in these networks. However, we did not observe noticeable drops in performance for the other dense networks, which suggests that OVNS is able to generalize beyond the specific generative model. Also, it should be mentioned that for OBMA, we needed to considerably relax the number of iterations during each tabu search (see Appendix \ref{app:hyperparam-selection}). This will likely affect overall performance negatively. The fact that this relaxation was necessary in MDP networks which were used in \citet{zhou2017opposition} is indicative of the fact that our python implementation of OBMA is significantly slower than the original C++ implementation \cite{zhou2017opposition}. However, this bottle neck on performance applies to OVNS as well – i.e. we expect that implementing OVNS in high performance languages such as Julia, C, or C++ can produce considerable speedups.

\subsection{Directions for future research}
 Benchmarking on even larger networks is a potential future research direction, and it would reveal more differences in the scaling of the heuristics.
 Other future directions include testing multiple simultaneous moves during the neighborhood search phase, implementing efficient parallelization, and guiding the search using memory structures such as tabu lists. In their survey, \citet{hussain2019metaheuristic} also foresee a promising avenue of research for a new adaptive techniques that allow heuristics to self-tune their parameter values based on objective function values during execution. In OVNS, we already introduce the $q$ parameter for min-thresholding the weights of the input network. In addition, parameters such as $p_{step}$ size, and neighborhood search mode can be taken advantage of for fine-tuning and self-adapting the search during execution. In OVNS, we combine two heuristics (drop and BVNS) into a hybrid heuristic that surpasses both of their individual performance. Similar hybridization techniques could be further explored with the aim of producing new heuristics with state-of-the-art performance. For instance, OBMA constructs the opposite solution candidate by picking nodes uniformly in random from the rest of the network's nodes. Further work could study the potential performance gains if stratified sampling proportional to degrees would be adopted in OBMA.

\appendix
\section{Appendix}
Source code for the implementation of OVNS, OBMA and BVNS is available at \url{https://github.com/Decitizen/OVNS}

\begin{table}[!htb]
    \centering
    \begin{tabular}{lrrrrr}
    \toprule
        {}        & $p_{step}$ & search & shake & $q$ \\
        \midrule
        BVNS     &  1        & \textit{first} & \textit{uniform}* & 1.00* \\
        OVNS     &  $\lfloor k/10 \rfloor$    & \textit{first} & \textit{preferential} & 1.00 \\
        \bottomrule 
    \end{tabular}

    \caption{Table listing the parametrizations for BVNS and OVNS for the benchmarking runs. Additionally, OVNS uses the drop heuristic for initialization while BVNS uses random initialization (best out of 1000 draws). (* these parameter values don't exist in BVNS, but corresponding parameter values in OVNS have been added for reference.)}
    \label{tab:O-VNS_param}
\end{table}

\subsection{Data sources}
\label{app:data-sources}
Table \ref{tab:my_label} describes used social networks. We obtained data sets from two sources, i) 5 social networks from the open access data repository \textit{networkrepository.com} \cite{nr}, as well as 33 Twitter retweet networks introduced in \cite{chen2021polarization}. For networkrepository.com we narrowed the search to weighted, multi-edged or temporal social networks with a focus on online human communication networks such as email correspondence networks. We limited the size of the networks to range $[10^2,10^5]$, and reduced all temporal or multilayer networks to static undirected single layer networks. For temporal networks, we derived weights by aggregating temporal edges between nodes. For MDP networks we used the subset of mdplib 2.0 benchmarking set \cite{marti2021mdplib}. To allow easy comparisons, we used the same MDG-a 20-40 and MDG-c 20-40 instances that Zhou et al. used in their OBMA paper \cite{zhou2017opposition}.

\begin{table}[]
\footnotesize
    \centering
\begin{tabular}{lr}
\toprule
network               &     $n$ \\
\midrule
\multicolumn{2}{c}{Twitter networks \cite{chen2021polarization} ($N=33$)} \\
\midrule
Centre 1              &   1600 \\
Centre 2              &   1488 \\
Centre 3              &   1030 \\
Climate 1             &  15,061 \\
Climate 2             &   6951 \\
Climate 3             &   6567 \\
Economicpolicy 1      &   3328 \\
Economicpolicy 2      &   2734 \\
Economicpolicy 3      &   2828 \\
Education 1           &   7563 \\
Education 2           &   4312 \\
Education 3           &   4323 \\
Finns 1               &   1680 \\
Finns 2               &   1685 \\
Finns 3               &   1743 \\
Green 1               &   2225 \\
Green 2               &   1537 \\
Green 3               &   1757 \\
Immigration 1         &   3373 \\
Immigration 2         &   2040 \\
Immigration 3         &   2879 \\
Left 1                &   1226 \\
Left 2                &    783 \\
Left 3                &    531 \\
National 1            &   2703 \\
National 2            &   1466 \\
National 3            &    877 \\
Sdp 1                 &   2084 \\
Sdp 2                 &   1424 \\
Sdp 3                 &    732 \\
Socialsecurity 1      &   7629 \\
Socialsecurity 2      &   3816 \\
Socialsecurity 3      &   3597 \\
\midrule
\multicolumn{2}{c}{Network repository \cite{nr} ($N=5$)}  \\
\midrule
EMAIL-DNC-1 &    906 \\
EMAIL-ENRON-L*     &  33,696 \\
EMAIL-DNC-2           &   1891 \\
SOC-WIKI-ELEC         &   7118 \\
SOC-WIKI-VOTE         &    889 \\
\midrule
Dense transformed networks & ($N=5$) \\
\midrule
NBKT-Climate 1**           &  15,061 \\
NBKT-Education 1**       &   7563 \\
NBKT-Immigration 1**       &   3373 \\
NBKT-BBV1**                &  10,000 \\
NBKT-BBV2**                &  10,000 \\

\bottomrule
\end{tabular}
    \caption{Sparse and dense benchmarking networks and their sizes. Number of runs for each algorithm and each parameter setting was 20, time budget per run was 10 minutes. Exceptions: (*) 15 runs, 30 minutes, (**) 20 runs, 60 minutes.}
    \label{tab:my_label}
\end{table}

\subsection{Parameters in benchmarks}
\label{app:hyperparam-selection}
Table \ref{tab:O-VNS_param} shows the parameter settings for BVNS and OVNS heuristics. Setting $p_{step} = \frac{k}{10}$ makes the step size dependent on the problem difficulty and aims to allow rapid diversification also in cases where $k$ is large.

For OBMA, the standard parameter values described in \cite{zhou2017opposition} were used with the only difference being the number of iterations $n_{tabu}$ allowed for each round of tabu search neighborhood search. For the python implementation, the original default value reported in the paper of 50,000 was too high resulting in OBMA running over the given time limit. We relaxed this parameter value and determined it dynamically based on network size $N$, by using the following heuristic: $n_{tabu} = 1.5\cdot10^6 / N$. 

\subsection{Broader perspectives and ethical considerations}
Our work focuses on improving existing heuristic approaches for dense subgraph finding with structural insights derived from real-life social networks, and while our method offers a significant performance improvement in social networks, similar results can be achieved by committing more computational resources to the task.
Therefore, it is unlikely that our approach will have any disruptive or harmful impact on society caused by threats that were made possible by this research. However, despite the potential for many beneficial societal impacts related to social dynamics, epidemiology, law enforcement, managerial processes, urban planning, logistics, ecology, and biology, the identification of groups in social networks through the use of methods such as ours must be approached with caution and requires careful consideration of potential second and third order consequences in the given application context. 

When analyzing human social networks, we strongly encourage users of our approach to consider the right to privacy of the individuals represented in the network. It is crucial to properly anonymize network data to ensure that the privacy and data rights of identified individuals are not violated. We would also like to emphasize that especially in the context of experimental application in social networks and socially embedded systems, researchers should seek for informed consent whenever it is possible given the constraints of the application context. Both of these practices are of paramount importance in order to ensure that potential beneficial impacts of the research are maximized while minimizing any negative consequences.

\section{Acknowledgments} 
 We would like to thank both the Aalto Science-IT project for their computational resources as well as the Aalto SciComp for technical assistance and support. This work was supported by the Academy of Finland (grant numbers 320780, 320781, and 349366).
\bibliography{bib}

\begin{thebibliography}{35}
\providecommand{\natexlab}[1]{#1}

\bibitem[{Aringhieri and Cordone(2011)}]{aringhieri2011comparing}
Aringhieri, R.; and Cordone, R. 2011.
\newblock Comparing local search metaheuristics for the maximum diversity
  problem.
\newblock \emph{Journal of the Operational research Society}, 62(2): 266--280.

\bibitem[{Arrigo et~al.(2022)Arrigo, Higham, Noferini, and
  Wood}]{arrigo2022weighted}
Arrigo, F.; Higham, D.~J.; Noferini, V.; and Wood, R. 2022.
\newblock Weighted enumeration of nonbacktracking walks on weighted graphs.

\bibitem[{Asahiro et~al.(2000)Asahiro, Iwama, Tamaki, and
  Tokuyama}]{asahiro2000greedily}
Asahiro, Y.; Iwama, K.; Tamaki, H.; and Tokuyama, T. 2000.
\newblock Greedily Finding a Dense Subgraph.
\newblock \emph{Journal of Algorithms}, 34(2): 203--221.

\bibitem[{Barabási and Albert(1999)}]{barabasi1999emergence}
Barabási, A.-L.; and Albert, R. 1999.
\newblock Emergence of Scaling in Random Networks.
\newblock \emph{Science}, 286(5439): 509--512.

\bibitem[{Barrat et~al.(2004)Barrat, Barthelemy, Pastor-Satorras, and
  Vespignani}]{barrat2004architecture}
Barrat, A.; Barthelemy, M.; Pastor-Satorras, R.; and Vespignani, A. 2004.
\newblock The architecture of complex weighted networks.
\newblock \emph{Proceedings of the national academy of sciences}, 101(11):
  3747--3752.

\bibitem[{Barrat, Barth{\'e}lemy, and Vespignani(2004)}]{barrat2004weighted}
Barrat, A.; Barth{\'e}lemy, M.; and Vespignani, A. 2004.
\newblock Weighted evolving networks: coupling topology and weight dynamics.
\newblock \emph{Physical review letters}, 92(22): 228701.

\bibitem[{Benigni, Joseph, and Carley(2019)}]{benigni2019bot}
Benigni, M.~C.; Joseph, K.; and Carley, K.~M. 2019.
\newblock Bot-ivistm: assessing information manipulation in social media using
  network analytics.
\newblock In \emph{Emerging research challenges and opportunities in
  computational social network analysis and mining}, 19--42. Springer.

\bibitem[{Billionnet(2005)}]{billionnet2005different}
Billionnet, A. 2005.
\newblock Different formulations for solving the heaviest k-subgraph problem.
\newblock \emph{INFOR: Information Systems and Operational Research}, 43(3):
  171--186.

\bibitem[{Blum and Roli(2003)}]{blum2003metaheuristics}
Blum, C.; and Roli, A. 2003.
\newblock Metaheuristics in Combinatorial Optimization: Overview and Conceptual
  Comparison.
\newblock \emph{ACM Comput. Surv.}, 35(3): 268–308.

\bibitem[{Brimberg et~al.(2009)Brimberg, Mladenović, Urošević, and
  Ngai}]{brimberg2009variable}
Brimberg, J.; Mladenović, N.; Urošević, D.; and Ngai, E. 2009.
\newblock Variable neighborhood search for the heaviest k-subgraph.
\newblock \emph{Computers \& Operations Research}, 36(11): 2885--2891.

\bibitem[{Chen and Saad(2010)}]{chen2010dense}
Chen, J.; and Saad, Y. 2010.
\newblock Dense subgraph extraction with application to community detection.
\newblock \emph{IEEE Transactions on knowledge and data engineering}, 24(7):
  1216--1230.

\bibitem[{Chen et~al.(2021)Chen, Salloum, Gronow, Yl{\"a}-Anttila, and
  Kivel{\"a}}]{chen2021polarization}
Chen, T. H.~Y.; Salloum, A.; Gronow, A.; Yl{\"a}-Anttila, T.; and Kivel{\"a},
  M. 2021.
\newblock Polarization of climate politics results from partisan sorting:
  Evidence from Finnish Twittersphere.
\newblock \emph{Global Environmental Change}, 71: 102348.

\bibitem[{Clauset, Shalizi, and Newman(2009)}]{clauset2009power}
Clauset, A.; Shalizi, C.~R.; and Newman, M.~E. 2009.
\newblock Power-law distributions in empirical data.
\newblock \emph{SIAM review}, 51(4): 661--703.

\bibitem[{Corneil and Perl(1984)}]{corneil1984clustering}
Corneil, D.; and Perl, Y. 1984.
\newblock Clustering and domination in perfect graphs.
\newblock \emph{Discrete Applied Mathematics}, 9(1): 27--39.

\bibitem[{Dokeroglu et~al.(2019)Dokeroglu, Sevinc, Kucukyilmaz, and
  Cosar}]{dokeroglu2019survey}
Dokeroglu, T.; Sevinc, E.; Kucukyilmaz, T.; and Cosar, A. 2019.
\newblock A survey on new generation metaheuristic algorithms.
\newblock \emph{Computers \& Industrial Engineering}, 137: 106040.

\bibitem[{Dong et~al.(2021)Dong, Chen, Tricco, Li, and Hu}]{dong2021hunting}
Dong, Z.; Chen, Y.; Tricco, T.~S.; Li, C.; and Hu, T. 2021.
\newblock Hunting for vital nodes in complex networks using local information.
\newblock \emph{Scientific Reports}, 11(1): 1--13.

\bibitem[{Dourisboure, Geraci, and
  Pellegrini(2007)}]{dourisboure2007extraction}
Dourisboure, Y.; Geraci, F.; and Pellegrini, M. 2007.
\newblock Extraction and classification of dense communities in the web.
\newblock In \emph{Proceedings of the 16th international conference on World
  Wide Web}, 461--470.

\bibitem[{Hansen et~al.(2019)Hansen, Mladenovi{\'{c}}, Brimberg, and
  P{\'e}rez}]{hansen2019variable}
Hansen, P.; Mladenovi{\'{c}}, N.; Brimberg, J.; and P{\'e}rez, J. A.~M. 2019.
\newblock Variable Neighborhood Search.
\newblock In Gendreau, M.; and Potvin, J.-Y., eds., \emph{Handbook of
  Metaheuristics}, 57--97. Springer International.

\bibitem[{Holme(2019)}]{holme2019rare}
Holme, P. 2019.
\newblock Rare and everywhere: Perspectives on scale-free networks.
\newblock \emph{Nature communications}, 10(1): 1--3.

\bibitem[{Hussain et~al.(2019)Hussain, Salleh, Cheng, and
  Shi}]{hussain2019metaheuristic}
Hussain, K.; Salleh, M. N.~M.; Cheng, S.; and Shi, Y. 2019.
\newblock Metaheuristic research: a comprehensive survey.
\newblock \emph{Artificial Intelligence Review}, 52(4): 2191--2233.

\bibitem[{Karp(1972)}]{karp1972reducibility}
Karp, R.~M. 1972.
\newblock Reducibility among combinatorial problems.
\newblock In \emph{Complexity of computer computations}, 85--103. Springer.

\bibitem[{Letsios et~al.(2016)Letsios, Balalau, Danisch, Orsini, and
  Sozio}]{letsios2016finding}
Letsios, M.; Balalau, O.~D.; Danisch, M.; Orsini, E.; and Sozio, M. 2016.
\newblock Finding heaviest k-subgraphs and events in social media.
\newblock In \emph{2016 IEEE 16th International Conference on Data Mining
  Workshops (ICDMW)}, 113--120. IEEE.

\bibitem[{L{\"u} et~al.(2016)L{\"u}, Chen, Ren, Zhang, Zhang, and
  Zhou}]{lu2016vital}
L{\"u}, L.; Chen, D.; Ren, X.-L.; Zhang, Q.-M.; Zhang, Y.-C.; and Zhou, T.
  2016.
\newblock Vital nodes identification in complex networks.
\newblock \emph{Physics Reports}, 650: 1--63.

\bibitem[{Mart{\'\i} et~al.(2013)Mart{\'\i}, Gallego, Duarte, and
  Pardo}]{marti2013heuristics}
Mart{\'\i}, R.; Gallego, M.; Duarte, A.; and Pardo, E.~G. 2013.
\newblock Heuristics and metaheuristics for the maximum diversity problem.
\newblock \emph{Journal of Heuristics}, 19(4): 591--615.

\bibitem[{Martí et~al.(2021)Martí, Duarte, Martínez-Gavara, and
  Sánchez-Oro}]{marti2021mdplib}
Martí, R.; Duarte, A.; Martínez-Gavara, A.; and Sánchez-Oro, J. 2021.
\newblock The MDPLIB 2.0 Library of Benchmark Instances for Diversity Problems.

\bibitem[{Martí et~al.(2022)Martí, Martínez-Gavara, Pérez-Peló, and
  Sánchez-Oro}]{MARTI2022795}
Martí, R.; Martínez-Gavara, A.; Pérez-Peló, S.; and Sánchez-Oro, J. 2022.
\newblock A review on discrete diversity and dispersion maximization from an OR
  perspective.
\newblock \emph{European Journal of Operational Research}, 299(3): 795--813.

\bibitem[{Mladenovi{\'c} and Hansen(1997)}]{mladenovic1997variable}
Mladenovi{\'c}, N.; and Hansen, P. 1997.
\newblock Variable neighborhood search.
\newblock \emph{Computers \& operations research}, 24(11): 1097--1100.

\bibitem[{Newman and Park(2003)}]{newman2003social}
Newman, M.~E.; and Park, J. 2003.
\newblock Why social networks are different from other types of networks.
\newblock \emph{Physical review E}, 68(3): 036122.

\bibitem[{Onnela et~al.(2007)Onnela, Saram{\"a}ki, Hyv{\"o}nen, Szab{\'o},
  Lazer, Kaski, Kert{\'e}sz, and Barab{\'a}si}]{onnela2007structure}
Onnela, J.-P.; Saram{\"a}ki, J.; Hyv{\"o}nen, J.; Szab{\'o}, G.; Lazer, D.;
  Kaski, K.; Kert{\'e}sz, J.; and Barab{\'a}si, A.-L. 2007.
\newblock Structure and tie strengths in mobile communication networks.
\newblock \emph{Proceedings of the national academy of sciences}, 104(18):
  7332--7336.

\bibitem[{Parreño, Álvarez Valdés, and Martí(2021)}]{PARRENO2021515}
Parreño, F.; Álvarez Valdés, R.; and Martí, R. 2021.
\newblock Measuring diversity. A review and an empirical analysis.
\newblock \emph{European Journal of Operational Research}, 289(2): 515--532.

\bibitem[{Rossi and Ahmed(2015)}]{nr}
Rossi, R.~A.; and Ahmed, N.~K. 2015.
\newblock The Network Data Repository with Interactive Graph Analytics and
  Visualization.
\newblock In \emph{AAAI}.

\bibitem[{S{\"o}rensen, Sevaux, and Glover(2018)}]{sorensen2018history}
S{\"o}rensen, K.; Sevaux, M.; and Glover, F. 2018.
\newblock A history of metaheuristics.
\newblock In \emph{Handbook of heuristics}, 791--808. Springer.

\bibitem[{Toivonen et~al.(2009)Toivonen, Kovanen, Kivel{\"a}, Onnela,
  Saram{\"a}ki, and Kaski}]{toivonen2009comparative}
Toivonen, R.; Kovanen, L.; Kivel{\"a}, M.; Onnela, J.-P.; Saram{\"a}ki, J.; and
  Kaski, K. 2009.
\newblock A comparative study of social network models: Network evolution
  models and nodal attribute models.
\newblock \emph{Social networks}, 31(4): 240--254.

\bibitem[{Ugander et~al.(2013)Ugander, Karrer, Backstrom, and
  Kleinberg}]{ugander2013graph}
Ugander, J.; Karrer, B.; Backstrom, L.; and Kleinberg, J. 2013.
\newblock Graph cluster randomization: Network exposure to multiple universes.
\newblock In \emph{Proceedings of the 19th ACM SIGKDD international conference
  on Knowledge discovery and data mining}, 329--337.

\bibitem[{Zhou, Hao, and Duval(2017)}]{zhou2017opposition}
Zhou, Y.; Hao, J.-K.; and Duval, B. 2017.
\newblock Opposition-based memetic search for the maximum diversity problem.
\newblock \emph{IEEE Transactions on Evolutionary Computation}, 21(5):
  731--745.

\end{thebibliography}

\end{document}